\documentclass{article}
\usepackage[T1]{fontenc} 
\usepackage[utf8]{inputenc} 
\usepackage{ismir,amsmath,cite,url}
\usepackage{graphicx}
\usepackage{color}
\graphicspath{{figs/}}

\usepackage{array}
\newcommand{\PreserveBackslash}[1]{\let\temp=\\#1\let\\=\temp}
\newcolumntype{C}[1]{>{\PreserveBackslash\centering}p{#1}}
\newcolumntype{R}[1]{>{\PreserveBackslash\raggedleft}p{#1}}
\newcolumntype{L}[1]{>{\PreserveBackslash\raggedright}p{#1}}

\usepackage{algorithm}
\usepackage{algorithmic}
\usepackage{bbm}
\usepackage{lineno}

\title{BeatNet: CRNN and particle filtering for online joint beat downbeat and meter tracking}

\oneauthor
   {Mojtaba Heydari \hspace{1.7cm}   Frank Cwitkowitz \hspace{1.7cm}   Zhiyao Duan} {Department of Electrical and Computer Engineering \\
University of Rochester, 500 Wilson Blvd, Rochester, NY 14627, USA \\ {\tt \( \{ \)mheydari,fcwitkow\( \} \)@ur.rochester.edu \text{}  zhiyao.duan@rochester.edu}}

\def\authorname{M. Heydari, F. Cwitkowitz, and Z. Duan}

\begin{document}

\maketitle
\begin{abstract}
The online estimation of rhythmic information, such as beat positions, downbeat positions, and meter, is critical for many real-time music applications. Musical rhythm comprises complex hierarchical relationships across time, rendering its analysis intrinsically challenging and at times subjective. Furthermore, systems which attempt to estimate rhythmic information in real-time must be causal and must produce estimates quickly and efficiently. In this work, we introduce an online system for joint beat, downbeat, and meter tracking, which utilizes causal convolutional and recurrent layers, followed by a pair of sequential Monte Carlo particle filters applied during inference. The proposed system does not need to be primed with a time signature in order to perform downbeat tracking, and is instead able to estimate meter and adjust the predictions over time. Additionally, we propose an information gate strategy to significantly decrease the computational cost of particle filtering during the inference step, making the system much faster than previous sampling-based methods. Experiments on the GTZAN dataset, which is unseen during training, show that the system outperforms various online beat and downbeat tracking systems and achieves comparable performance to a baseline offline joint method.
\end{abstract}
\begin{figure*}[htbp]
 \centerline{
 \includegraphics[width=0.9\textwidth]{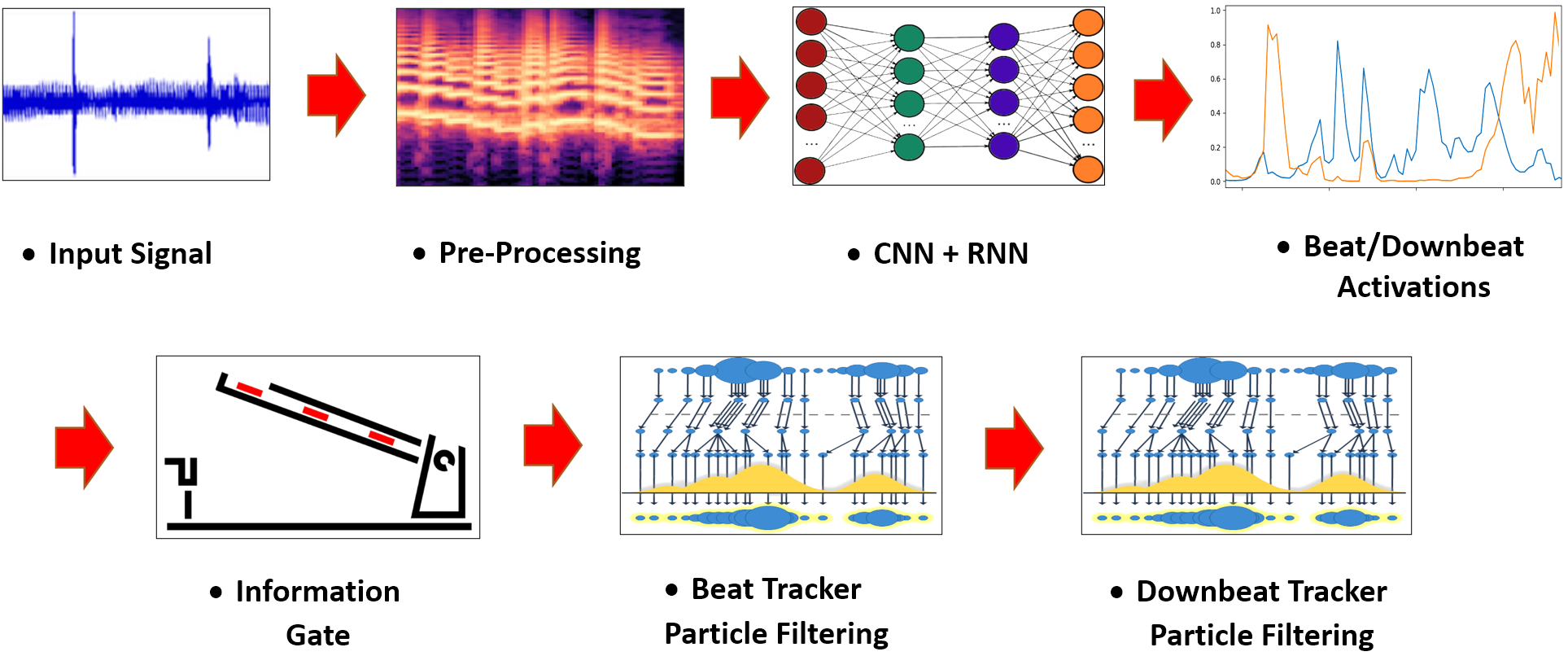}}
 \caption{Overview of the joint beat, downbeat, and meter tracking procedure using the proposed BeatNet model.}
 \label{fig1}
\end{figure*}

\section{Introduction}\label{sec:introduction}
Rhythm plays an essential role in nearly all musical endeavors, including listening to, playing, learning, or composing music. This is why the estimation of rhythmic information, such as beat positions, downbeat positions and meter has always been an important subject of study in the field of Music Information Retrieval (MIR). Depending on the requirements and constraints imposed by the application at hand, these estimation tasks can either be performed in an offline or online fashion. Offline approaches are typically non-causal, meaning that they make predictions for a given time using data or features associated with a future time. These approaches are suitable for applications such as music transcription, music search and indexing, and musicological analysis. Online approaches are causal, meaning that they operate using only past and present features. These are typically desirable for human-computer interaction (HCI) systems, which must make immediate predictions, like real-time music accompaniment systems.

Many offline methods have been proposed for beat tracking \cite{ellis,davis,gouyon}. Most of them are unsupervised and attempt to utilize low-level features like onset strengths with some inference model to estimate beat positions within a music piece. However, with the growing success of deep learning, supervised beat tracking methods have become more prominent. Böck et al. \cite{Bock:2} employed Recurrent Neural Networks (RNNs) to estimate beat positions; Various other neural network structures have also been proposed for onset detection and beat tracking \cite{davis:2,Bock:3}. 

Some methods have also been proposed for online beat tracking. However, many of them, e.g., \cite{davis:3,Gkiokas,Brossier,Bock:2,mottaghi}, feed a sliding window of data into an offline model to estimate beat positions within upcoming frames. The sliding window strategy has several major drawbacks, including the discontinuity of beat predictions and the need for priming for predictions in the first window, which causes a delay \cite{Oliveria}. Some other approaches involve inferring beat positions in real time using multi agent models \cite{goto,goto:2,goto:3,Oliveria}, which initialize a set of agents with various hypotheses that try to validate their respective hypotheses based on observations across time.

The task of downbeat tracking is often considered to be more difficult than beat tracking. This is because a deeper understanding of rhythmic structure in music is required to be able to differentiate between beats and downbeats. Making matters worse, at the signal level, these two events have very similar characteristics. For instance, downbeats are not necessarily associated with stronger signal energy, nor do they necessarily feature a distinct percussive profile. Moreover, both beats and downbeats are likely to be the intersection of melodic and harmonic changes. These factors can make it challenging, and in some cases subjective, to distinguish between the two rhythmic events. For instance, for a 4/4 music piece with kick drum events on the first and third beats, it is hard to distinguish downbeats and determine whether the time signature is 4/4 or 2/4.  

There has been some previous work on offline downbeat tracking, both as an isolated task and within a joint beat and downbeat tracking framework. Durand et al.~\cite{durand:1,durand:2,durand:3} used some combinations of features and CNN structures to obtain downbeats. Giorgi et al. \cite{Giorgi} proposed tempo-invariant convolutional filters for downbeat tracking. Peeters and Papadopoulos~\cite{Petters} performed joint beat and downbeat tracking by decoding hidden states using the Viterbi algorithm. Böck et al.~\cite{Bock:1} and Krebs et al.~\cite{krebs} employed an RNN structure for joint beat and downbeat tracking and only downbeat tracking using beat synchronous features, respectively. Furthermore, some recent works investigate Convolutional Recurrent Neural Network (CRNN) structures for beat and downbeat tracking. Fuentes et al.~\cite{Fuentes:01} showed that CRNN structures outperform RNNs in downbeat tracking when taking the input observations over a tatum grid. Cheng et al.~\cite{Cheng} found that CRNN structures with larger receptive fields outperform other downbeat tracking models. Böck and Davies.~\cite{bock:07} used a CNN and Temporal Convolutional Network (TCN) structure to improve the performance of their offline beat and downbeat tracking model, and also performed data augmentation to expose the neural network to more tempi.

The task of online downbeat tracking has received considerably less attention. Goto and Muranoka \cite{goto:2} introduced an unsupervised model which leverages a measure inference stage for detecting chord changes. In \cite{Bock:5}, the same beat tracking neural network with forward algorithm from \cite{Bock:1,Bock:2} is paired with \cite{krebs} to estimate downbeats and other rhythmic patterns by extracting percussive and harmonic beat-synchronous features. It is important to note that this method must be primed with a known time signature and all possible rhythmic pattern choices. Liang~\cite{Liang} proposed an online downbeat tracking method which feeds a sliding window of data to an offline model~\cite{durand:3}. This method is vulnerable to the sliding window strategy drawbacks described above.


Particle filtering is advantageous for two main reasons when it comes to online processing. The first reason is that it does not require future data. Popular maximum a posteriori (MAP) algorithms like the Viterbi algorithm and maximizer of the posterior marginals (MPM) smoothing algorithms, e.g. forward-backward, are not applicable to online processing. The second reason is that, among the filtering methods which are causal, particle filtering is a general (non-parametric) approach which can be utilized to decode any unknown distribution. However, most music rhythmic analysis approaches that utilize particle filtering, e.g., \cite{Hainsworth,Hainsworth:1,Cemgil,Krebs:3}, are classical and do not incorporate neural networks. Alternatively, in our previous work \cite{Heydari}, we utilized a particle filtering inference model to infer beat positions using the activations produced by an RNN in an online fashion, but that approach does not attempt to estimate downbeats nor meter.

In this paper, we propose BeatNet, a novel online system for joint beat, downbeat, and meter tracking. The system produces beat and downbeat activations using a CNN and RNN combination, and performs inference using two particle filtering stages. The beat tracking stage outperforms state-of-the-art online beat tracking methods. The other stage simultaneously infers downbeats and time signature 
and achieves comparable results to state-of-the-art offline downbeat tracking models that require the time signature as input.
In contrast, BeatNet 
actively monitors tempo and time signature changes over time. Finally, we introduce an information gate mechanism in the inference module to speed up the inference significantly, making our method suitable for many real-time applications.

\section{Method}
\label{sec:method}
In this section, we describe BeatNet, our online system for joint beat, downbeat, and meter tracking, illustrated in Figure 1. BeatNet consists of a causal neural network stage for producing activations and a particle filtering stage for inference. The neural network comprises convolutional, recurrent and fully connected layers as described in section 2.2 which compute beat and downbeat activations for each frame of audio. The activations are fed to a two-stage particle filtering module to infer beat and downbeat positions and to estimate meter. The code for the BeatNet model is open-source\footnote{https://github.com/mjhydri/BeatNet}, along with video demos and further documentation.

\subsection{Feature Representation}
The input of the network module is a sequence of filterbank magnitude responses, each of which corresponds to one audio frame. Specifically, short-time Fourier transform (STFT) with a Hann window of the length of 93 ms and hop size of 46 ms is applied to the audio signal to compute the log-amplitude magnitude spectrogram. Then a logarithmically spaced filterbank ranging from 30 Hz to 17 kHz with 24 bands per octave is applied to yield a 136-d filterbank response. The first-order temporal difference of this response is also calculated and concatenated, resulting in a 272-d filterbank response vector for each frame.

We also experimented with alternative feature representations, including the 329-d hand-crafted feature set from \cite{durand:1}, which comprises chroma features, onset strengths, low-frequency spectral features, and melodic constant-Q spectral features. The motivation for this feature set is to aggregate the harmonic, percussive, bass, and melodic content of the music. However the 272-d filterbank response feature set described above achieved notably better performance than these hand-crafted features, and was thus chosen for subsequent experiments.

\subsection{Network Architecture}
Following the common design of other similar works, we employ a convolutional-recurrent neural network (CRNN) architecture, illustrated in Figure 2, to process the input features in order to obtain beat and downbeat activations. Ideally, the convolution models relationships along the frequency axis, and the unidirectional recurrence models long-term relationships across time in a causal fashion.

The input features are fed into a 1D convolutional layer with 2 filters of kernel size 10, followed by ReLU activation. The two filter responses are max pooled with kernel size 2 along frequency and then concatenated into a single feature embedding for each frame.
Then, a fully-connected layer with 150 neurons reduces the dimensionality of the embedding, and feeds it through two subsequent unidirectional Long Short-Term Memory (LSTM) layers, each with a hidden size of 150. The embedding is then fed through a final fully-connected layer and a softmax operation to obtain three activations which represent beat, downbeat, and non-beat, respectively. Note that due to the softmax function, the final activations for each class always sum to one.


\begin{figure}[htbp]
 \centerline{
 \includegraphics[width=0.9\columnwidth]{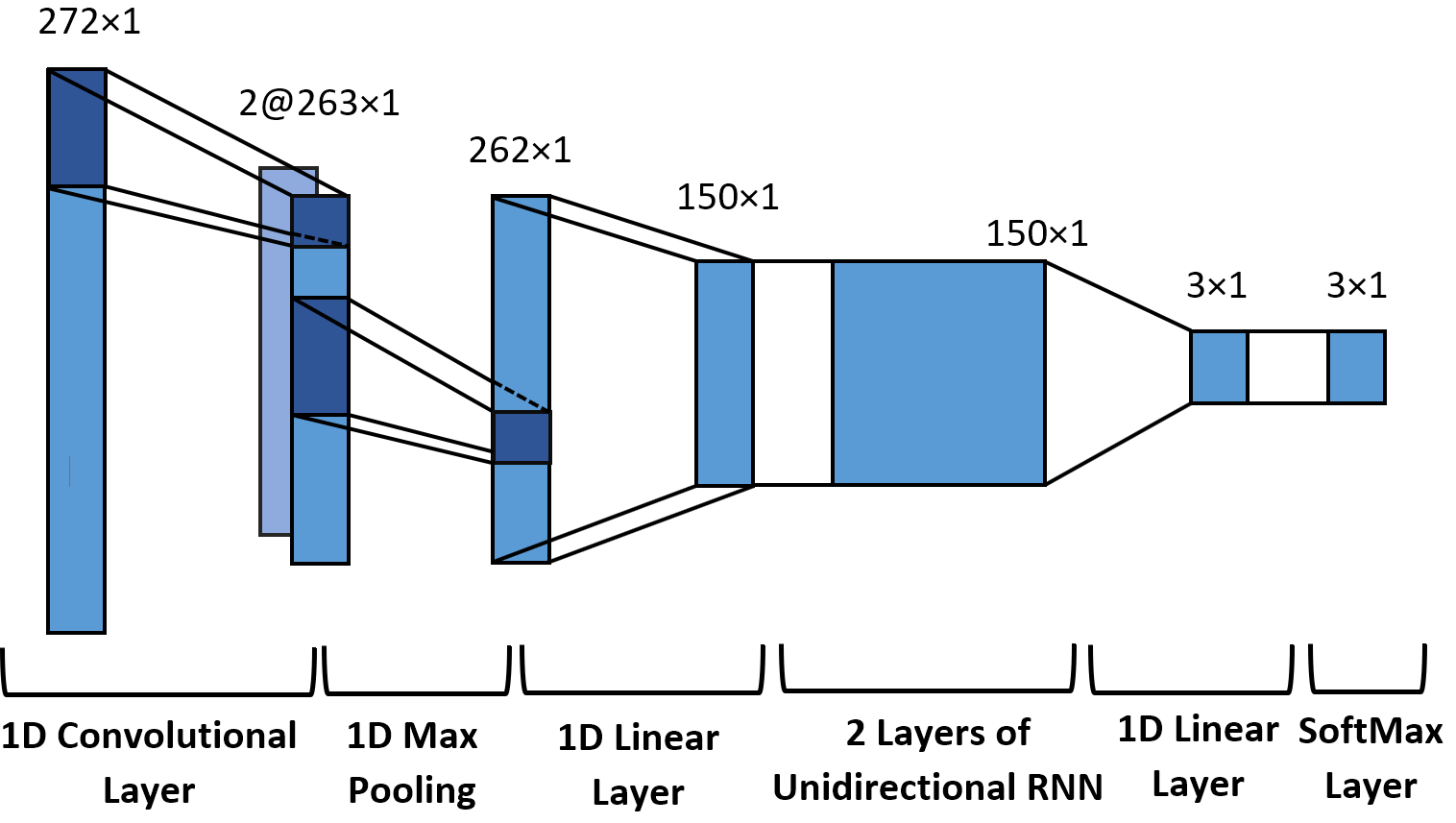}}
 \caption{Proposed CRNN architecture for processing input features and computing beat and downbeat activations.}
 \label{fig2}
\end{figure}

\subsection{Particle Filtering Inference}
In this section, we discuss the two-stage online Monte Carlo particle filtering inference module, which generates the beat and downbeat predictions. Sequential Monte Carlo particle filtering is a sampling-based model which iteratively estimates any unknown distribution $p(x)$ by gathering a large number of independent samples from an arbitrary proposal distribution. The unknown distribution of interest in our case, up to the $K$-th frame, is the following posterior $p(x_{1:K} | y_{1:K})$\ of underlying beat or downbeat positions $x_{1:K}$ conditioned on beat observations $y_{1:K}$. It can be inferred according to the key equations below. For more detailed information, please refer to our previous work \cite{Heydari}.
\begin{align}
    p(x) 
    &= \lim_{N \to \infty} \sum^N_{i=1} \frac{\omega^{(i)}}{\sum^N_{i=1} \omega^{(i)}} \delta \left(x - x^{(i)}\right),
\end{align}
\begin{equation}
    p\left(x_{1:K}^{(i)} | y_{1:K}\right) \propto \prod_{k=1}^K p\left(y_k|x_k^{(i)}\right) p\left(x_k^{(i)}|x_{k-1}^{(i)}\right),
\end{equation}
\begin{align}
    \omega_k^{(i)} = p \left( y_k|x_k^{(i)} \right) \omega_{k-1}^{(i)},
\end{align}
where $\omega^{(i)}$ is the importance weight of particle $i$, and $\delta(\cdot)$ is the Dirac function. Eq. (1) describes the estimation of $p(x)$ using a large number of particles $(N \rightarrow \infty )$ and their importance weights. Eq. (2) is a dynamic model which updates the posterior of each frame $k$ using the transition (motion) and observation (correction) probabilities. Eq. (3) describes a recursive process to update the importance weights using the current observation and the importance weights of the previous step.

\subsubsection{State spaces, transition and observation models}
We use a cascade of two sequential Monte Carlo particle filters, one for beat tracking, and the other for downbeat and meter tracking. The state space and transition model of the beat estimator are similar to \cite{Krebs:2}. The beat state space is a type of 2D bar pointer model and its transition for the phase (horizontal) and the tempo (vertical) of the frame are described in Eqs. (4) and (5), respectively. The phase of frame $k$ within a beat interval and the tempo at frame $k$ are respectively denoted by $\phi_{b,k}$ and $\dot\phi_{b,k}$. A constant $\lambda_{b}$ influences the intensity of potential jumps across the tempo axis.

We propose a new beat observation model in Eq. (6), where ${x_{b,k}}$ and ${y_{b,k}}$ are the beat state and beat observations at frame k. For non-beat states we allocate a small likelihood as $\gamma=0.03$ instead of using the non-beat activation output from the neural network. For beat frames, since downbeats can also be considered beats, we assess the maximum of the beat and downbeat activations. If the maximum exceeds a certain threshold, i.e., $T=0.4$, then it is set as the likelihood; Otherwise, $\gamma$ is used. When $\gamma$ is used, we also bypass the costly re-sampling step in the beat particle filtering module. Therefore, the threshold serves as an \emph{information gate}, through which the computational cost is significantly reduced. 
\begin{equation}
\phi_{b,k} = (\phi_{b,k-1} + \dot{\phi}_{b,k-1}) \mod (\phi_b^{max}+1),
\end{equation}
\begin{equation}
    p(\dot{\phi}_{b,k} | \dot{\phi}_{b,k-1}) =
      \left\{
        \begin{array}{ll}
            \exp{\left(-\lambda_{b} \left| \frac{\dot{\phi}_{b,k}}{\dot{\phi}_{b,k-1}} \right|\right)} & \hspace{0.25em} \text{if } \phi_{b,k}=0\\
            \mathbbm{1}(\dot{\phi}_{b,k} = \dot{\phi}_{b,k-1}) & \hspace{0.25em} \text{if } \phi_{b,k}>0 
        \end{array}
      \right. ,
\end{equation}
\begin{equation}
    p(y_{b,k} | x_{b,k}) =
      \left\{
        \begin{array}{ll}
            \max(b_k, d_k) & \quad \text{if } \phi_{b,k}=0\text{ and} \\
            {} & \quad \max(b_k, d_k) \geq T \\
            \gamma & \quad \text{otherwise}
        \end{array}
      \right. ,
\end{equation}

The second particle filter detects downbeats and the time signature jointly. The state space is similar to that of beat tracking. However, here we introduce $\dot\phi_{d,k}$ corresponding to the meter, i.e., $\dot\phi_{d,k} \in {2,3,...,\dot\phi_{d}^{max}}$, and $\phi_{d,k}$ to describe the phase of the beat within the bar interval, i.e. $\phi_{d,k} \in {0,1,2,...,\phi_{d}^{max}}$. Eqs. (7) and (8) describe the phase and meter transition models. We only let meter change at the states belonging to the downbeat area i.e. $\phi_{d,k}=0$, and $\lambda_{d}$ is a constant parameter that decides what percent of the particles jump to other meters at the downbeat states. Also, in Eq. (9) we define the observation model used in the downbeat particle filter. The first states within the bar (downbeat area) take the downbeat activation and the rest of them (beat states) take the beat activation. Note that as the second particle filter operates less often, i.e., only when a beat is detected, no information gate is needed here.
\begin{equation}
    \phi_{d,k}=(\phi_{d,k-1}+\dot{\phi}_{d,k-1}) \mod (\phi_d^{max}+1),
\end{equation}
\begin{equation}
    p(\dot{\phi}_{d,k} | \dot{\phi}_{d,k-1}) =
      \left\{
        \begin{array}{ll}
            \lambda_{d} & \hspace{0.25em} \text{if } \phi_{d,k}=0 \text{ and}\\ & \hspace{0.25em} \dot{\phi}_{d,k} \neq \dot{\phi}_{d,k-1}\\
            1-\lambda_{d} & \hspace{0.25em} \text{if } \phi_{d,k}=0 \text{ and ,}\\ & \hspace{0.25em} \dot{\phi}_{d,k} = \dot{\phi}_{d,k-1}\\
            \mathbbm{1}(\dot{\phi}_{d,k} = \dot{\phi}_{d,k-1}) & \hspace{0.25em} \text{if } \phi_{d,k} > 0
        \end{array}
      \right.
\end{equation}
\begin{equation}
    p(y_{d,k} | x_{d,k}) =
      \left\{
        \begin{array}{ll}
            d_k & \quad \text{if } \phi_{d,k}=0\\
            b_k & \quad \text{if } \phi_{d,k}>0
        \end{array}
      \right. ,
\end{equation}

\subsubsection{Inference process}
Algorithm 1 describes the inference process in detail. Particles are initialized randomly for both inference modules by sampling from a uniform distribution within their state space. By proceeding to a new frame, particles within the beat state space are transferred to the new positions by sampling from the transition model, and new importance weights are then calculated and normalized. If the activations of the frame satisfy the information gate condition, the re-sampling process is invoked for all particles; Otherwise, the re-sampling step is skipped as it is likely a non-beat frame. 
Afterwards, if the median of the particles is within the tolerance window $T_{w}$ of a beat area and the time of the current frame is longer enough than the last detected beat considering the estimated tempo, the frame is classified as a beat frame. A similar process follows for the downbeat and meter inference module. 

\begin{figure}[htbp]
 \centerline{
 \includegraphics[width=0.9\columnwidth]{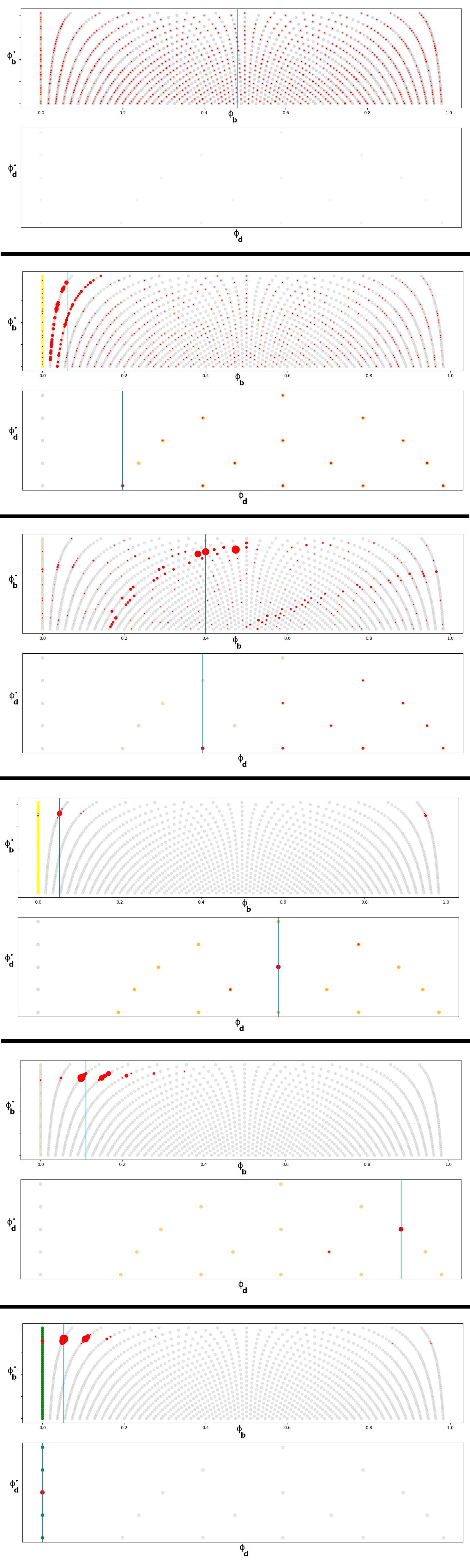}}
 \caption{Inference example, detailed in Section 2.3.2.}
 \label{fig2}
\end{figure}

\begin{algorithm}
    \caption{Joint Inference Procedure}
    \begin{algorithmic}
    \STATE beats, downbeats, meters = [], [], []
    \vspace{1mm}
      \STATE Sample $(x_{b,0}^{(i)}) \sim \mathcal U({S_b} )$, $(x_{d,0}^{(j)}) \sim \mathcal U({S_d})$
      \STATE Set $w_{b,0}^{(i)} = \frac{1}{N_b}$, $w_{d,0}^{(j)} = \frac{1}{N_d}$
    \vspace{1mm}
    \FOR{$k = 1 \text{ to } K$}
        \STATE Sample $(x_{b,k}^{(i)}) \sim p(\phi_{b,k}^{(i)}|\phi_{b,k-1}^{(i)}),p(\dot{\phi}_{b,k}^{(i)}|\dot{\phi}_{b,k-1}^{(i)})$
        \STATE $\tilde\omega_{b,k}^{(i)} = \omega_{b,k-1}^{(i)} \times p(y_{b,k}|x_{b,k}^{(i)}) \;\;\; \forall i \in N_b$
        \STATE $\omega_{b,k}^{(i)}=\frac{\tilde\omega_{b,k}^{(i)}}{\sum\tilde\omega_{b,k}^{(i)}}  \;\;\; \forall i \in N_b $
        \IF{$\max(b_k, d_k) \geq T$}
            \STATE Resample $x_{b,k}^{(i)}$ according to $ \omega_{b,k}^{(i)}$
        \ENDIF
        \IF{$\text{ median}(\phi_{b,k}^{(i)})< T_{w} \textbf{ and } (k\Delta-beats[-1]) > 0.4 \text{ median}(\dot\phi_{b,k}^{(i)})$}
                \STATE Append (beats, $k\Delta$)
                \STATE Sample $(x_{d,k}^{(j)}) \sim p(\phi_{d,k}^{(j)}|\phi_{d,k-1}^{(j)}), p(\dot{\phi}_{d,k}^{(j)}|\dot{\phi}_{d,k-1}^{(j)})$
                \STATE $\tilde\omega_{d,k}^{(j)} = \omega_{d,k-1}^{(j)} \times p(y_{d,k}|x_{d,k}^{(j)}) \;\;\; \forall j \in N_d$
                \STATE $\omega_{d,k}^{(j)}=\frac{\tilde\omega_{d,k}^{(j)}}{\sum\tilde\omega_{d,k}^{(j)}}  \;\;\; \forall j \in N_d $
                \STATE Resample $x_{d,k}^{(j)}$ according to $\omega_{d,k}^{(j)}$
                \IF{mode$(\phi_{d,k}^{(j)})==0$}
                \STATE append (downbeats, $k\Delta$)
                \STATE append (meters, mode$(\dot\phi_{d,k}^{(j)})$)
                \ENDIF
        \ENDIF
    \ENDFOR
    \end{algorithmic}
\end{algorithm}

A visualization of the inference process is presented in Figure 3. Each pair of plots demonstrates one step of the inference procedure, where the top and the bottom plots show the beat and downbeat tracking process, respectively. In the first pair of plots, the beat particles are initialized randomly. In the second pair, the first beat is detected and the downbeat state particles are simultaneously initialized randomly. In the third pair, beat tracking particles have converged, but the downbeat particles have not yet converged. Here the downbeat clutter is located in the lowest row of the downbeat state space, which represents a six-beat time signature. The next few plot pairs illustrate convergence of both the beat and downbeat particles, producing an estimate of the tempo and beat phase (top plots), and the meter and bar phase (bottom plots).

\section{Experiments}
\subsection{Methodology}
In order to analyze the performance of BeatNet, we compare it to several publicly available online beat tracking methods, We additionally provide the online downbeat tracking performance of BeatNet for each of the experiments. Following standard evaluation practices, in this work, F-measure with a tolerance window of $T_{w}=\pm70\text{ }ms$ is used as the evaluation metric for all experiments.

We utilize all five datasets~\cite{Gouyon:2,Krebs:4,davis,Srinivasamurthy,Marchand,Tzanetakis,Clercq} described in Table 1 for training, validation, and testing, with different splits and arrangements for various experiments.
\begin{table}[t]
  \begin{center}
    \begin{tabular}{L{0.30\columnwidth}C{0.20\columnwidth}C{0.25\columnwidth}}
        \hline
        \textit{Dataset} & \textit{\# Files} & \textit{Total Length} \\
        \hline
        \small Ballroom \cite{Gouyon:2,Krebs:4} & \small 685 & \small 5 h 57 m \\
        \hline
        \small Beatles \cite{davis} & \small 180 & \small 8 h 9 m \\
        \hline
        \small Carnatic \cite{Srinivasamurthy} & \small 176 & \small 16 h 38 m \\
        \hline
        \small GTZAN \cite{Marchand,Tzanetakis} & \small 999 & \small 8 h 20 m \\
        \hline
        \small Rock Corpus \cite{Clercq} & \small 200 & \small 12 h 53 m \\
        \hline
    \end{tabular}
    \caption{Datasets used for training and testing.}
    \label{tab:results}
  \end{center}
\end{table}
In the first comparison, we evaluate BeatNet on the GTZAN dataset, which covers 10 different music genres and was unseen from training of all comparison methods.  
In order to demonstrate the generalization ability of our approach, we also experiment with two other comparison schema where we respectively set aside the Ballroom and Rock datasets during training and use them entirely for evaluation. Note that all of the supervised comparison methods included the Ballroom and Rock datasets in their training set, so we only compare BeatNet with unsupervised methods in these cases.




\subsection{Training Details}
For training the beat and downbeat activation neural network described in Section 2.2, all weights and biases are initialized randomly, and the network is trained using Adam optimizer with a learning rate of $5 \times 10^{-4}$ and a batch size of 200.
Since the number of non-beat frames within a music piece is typically much larger than the number of beat and downbeat frames, our objective function is chosen to be weighted cross entropy loss of the beat, downbeat, and non-beat, where the weights are inverse proportional to the frequency of occurrence of each type of frame.
Batches comprise 8-second long excerpts randomly sampled from each audio file available in the training set. Given that some datasets (e.g., Beatles) contain full songs and others (e.g., Ballroom) contain short excerpts of songs, we sample from longer audio files more often during the training Batch creation.
Training proceeds until the performance on the validation set has not increased over a span of 20 epochs for a given experiment.

\subsection{Results and Discussion}
\begin{table}[t]
  \begin{center}
    \begin{tabular}{L{0.3\columnwidth}C{0.235\columnwidth}C{0.235\columnwidth}}
        \hline
        \small\textit{Method} & \small\textit{F-Measure} & \small\textit{F-Measure} \\
         & \small\textit{Beats} & \small\textit{Downbeats} \\
        \hline\hline
        \multicolumn{3}{c}{\textbf{Comparison of Online Methods}} \\
        \hline
        \multicolumn{3}{c}{\textit{GTZAN Dataset}} \\
        \small Aubio \cite{Brossier} & \small 57.09 & \small --- \\
        \small\textbf{BeatNet} & \small \underline{75.44} & \small \underline{46.49} \\
        \small Böck ACF \cite{Bock:2} & \small 64.63 & \small --- \\
        \small Böck FF \cite{Bock:1,Bock:3} & \small 74.18 & \small --- \\
        \small DLB \cite{Heydari} & \small 73.77 & \small --- \\
        \small IBT \cite{Oliveria} & \small 68.99 & \small --- \\
        \hline
        \multicolumn{3}{c}{\textit{Ballroom Dataset}} \\
        \small Aubio \cite{Brossier} & \small 56.73 & \small --- \\
        \small\textbf{BeatNet} & \small \underline{77.41} & \small \underline{47.45} \\
        \small IBT \cite{Oliveria} & \small 70.79 & \small --- \\
        \hline
        \multicolumn{3}{c}{\textit{Rock Corpus Dataset}} \\
        \small Aubio \cite{Brossier} & \small 59.83 & \small --- \\
        \small\textbf{BeatNet} & \small \underline{73.13} & \small \underline{44.98} \\
        \small IBT \cite{Oliveria} & \small 68.55 & \small --- \\
        \hline\hline
        \multicolumn{3}{c}{\textbf{Comparison of Offline Methods}} \\
        \hline
        \multicolumn{3}{c}{\textit{GTZAN Dataset}} \\
        \small\textbf{BeatNet + DBN} & \small \underline{80.64} & \small \underline{54.07} \\
        \small Böck \cite{Bock:1} & \small 79.09 & \small 51.36 \\
        \hline
    \end{tabular}
    \caption{Comparison of BeatNet with other beat and downbeat tracking methods on various datasets.}
    \label{tab:results}
  \end{center}
\end{table}

The evaluation results of the proposed BeatNet model and comparison methods are presented in Table 2. All online comparison methods only perform beat tracking, and all except IBT~\cite{Oliveria} and Aubio~\cite{Brossier} are supervised methods using deep neural networks. We can see that the online beat tracking portion of BeatNet outperforms all comparison methods. The Böck FF ~\cite{Bock:1,Bock:3} and Don't Look Back (DLB) models~\cite{Heydari} achieve the next best performance. Böck FF uses the forward algorithm to estimate beats in a similar manner to the other online joint model described earlier~\cite{Bock:5}. Aside from the different neural network structures, the beat tracking inference processes of the DLB model~\cite{Heydari} and BeatNet are largely the same. The main difference is that the latter benefits from the information gate, which decreases the computational time drastically.

Additionally, we report the performance comparison with an offline joint beat and downbeat tracking model~\cite{Bock:1} on the GTZAN dataset. In this case, we replaced the particle filtering modules of BeatNet with the DBN used in~\cite{Bock:1} to directly compare neural network architectures in BeatNet and~\cite{Bock:1}. Same to~\cite{Bock:1}, we also provided the time signatures to the DBN. For~\cite{Bock:1}, we utilized the Madmom~\cite{bock:06} library, which is the official implementation of the paper.
Note that due to the existence of different GTZAN beat annotations, the reported offline results obtained by us differ from those of the original paper~\cite{Bock:1}. However, since we used the same annotations for all of the experiments, the offline comparison is valid.
As the table suggests, with the same DBN estimator, both neural networks yield similar results for beat tracking. However, for downbeat tracking, the BeatNet architecture yields marginally better performance. These results are interesting, since we are comparing a causal network to a non-causal network which leverages bidirectional recurrence. However, our network is larger and contains more parameters. 

The comparison between BeatNet (second row) and~\cite{Bock:1} (last row) is also interesting. BeatNet underperforms~\cite{Bock:1} by 3.65\% on beat tracking and by 4.9\% on downbeat tracking. However, it is noted that BeatNet is an online method and it does not require the time signature input, while~\cite{Bock:1} is offline method and it requires the time signature input.

One limitation of our model is that the performance of the downbeat tracker depends on the beat tracker. This means that if the beat tracker makes incorrect predictions, errors will carry through to the downbeat tracker. This is a common characteristic of cascade systems such as~\cite{Bock:5}. Another limitation is the high computation cost of sequential Monte Carlo particle filtering methods. This limitation has been partially addressed in our previous work~\cite{Heydari} by using efficient models, e.g.,~\cite{Krebs:2} in the inference stage. The information gate proposed in this paper further reduces the computational cost.

On a typical windows machine with AMD Ryzen 9 3900X CPU and 3.80 GHz clock, the processing time for the pre-processing stage and passing a frame through the neural network is 0.12 ms and 0.01 ms, respectively. These times are relatively insignificant, as the inference process takes more time. The inference process takes 5.23 and 8.87 seconds using 1000 and 1750 particles, respectively, to process a 30-sec long music excerpt. This is much faster than the previous sampling-based model~\cite{Heydari} which took 21.30 seconds using a 1000 particle setup. Larger numbers of particles lead to longer processing times with a roughly linear relationship. Hence, we reported these results using 1500 particles for the beat inference block and 250 for the downbeat inference block (1750 particles in total) to keep the process minimal.

\section{Conclusion}
We proposed BeatNet, a new online system for joint beat, downbeat, and meter tracking. The system incorporates a convolutional-recurrent neural network for generating beat and downbeat activations in each audio frame, and a two-stage particle filtering algorithm to estimate tempo, beats, downbeats, and musical meter. An information gate is added to the beat tracking particle filter to skip many re-sampling steps hence reduces the computational cost significantly. 
The system is compared to multiple online and offline  methods under various experimental conditions, and it achieves superior performance for both online beat and downbeat tracking.

\section{Acknowledgement}
This work has been partially supported by the National Science Foundation grants 1846184 and DGE-1922591.


\bibliography{BeatNet}

%
%
%
%
%

\end{document}